\documentclass[12pt]{revtex4-1}

\usepackage{amssymb,amsfonts,amsmath}

\usepackage[pdftex]{graphicx}
\usepackage{amsthm}
\usepackage{textcomp}
\usepackage{latexsym}

\usepackage[left]{lineno}

\theoremstyle{definition}

\theoremstyle{remark}

\newcommand{\mb}{\mathbf}

\begin{document}
\title{Robust exponential memory in Hopfield  networks}
\author{Christopher Hillar*}
\affiliation{Redwood Center for Theoretical Neuroscience, Berkeley, CA 94720}
\email{chillar@msri.org}
\author{Ngoc M. Tran*}
\affiliation{University of Texas, Austin, TX}
\email{tran.mai.ngoc@gmail.com}

%\linenumbers
% \blindtext

\begin{abstract}
The Hopfield recurrent neural network \cite{hopfield1982} is a classical auto-associative model of memory, in which collections of symmetrically-coupled McCulloch-Pitts \cite{mcculloch1943} neurons interact to perform emergent  computation \cite{hopfield1986computing}.  Although previous researchers have explored the potential of this network to solve combinatorial optimization problems \cite{hopfield1985, koch1986analog} and  store memories as attractors of its deterministic dynamics \cite{mceliece1987, amari1989characteristics}, a basic open problem is to design a family of Hopfield networks with a number of noise-tolerant memories that grows exponentially with neural population size. 
Here, we discover such networks by minimizing  \textit{probability flow} \cite{sohl2011}, a recently proposed objective for estimating parameters in discrete maximum entropy models.  By descending the gradient of the convex probability flow, our networks adapt synaptic weights to achieve robust exponential storage, even when presented with vanishingly small numbers of training patterns.  In addition to providing a new set of error-correcting codes that achieve Shannon's channel capacity bound \cite{Shan48a}, these networks also efficiently solve a variant of the hidden clique problem \cite{alon1998finding} in  computer science, opening new avenues for real-world applications of computational models originating from  biology.
\end{abstract}

\maketitle
\textbf{Introduction}. Discovered first by Pastur and Figotin \cite{figotin1977} as a simplified spin glass \cite{edwards1975} in statistical physics, the Hopfield model \cite{hopfield1982} is a recurrent network of $n$ linear threshold McCulloch-Pitts \cite{mcculloch1943} neurons that can store $n/(4 \ln n)$ binary patterns as distributed memories \cite{mceliece1987}.
%While several aspects of the Hopfield network appeared earlier (e.g., dynamics and  learning \cite{amari1972learning}), the model applies ideas of statistical mechanics to study basic neural computation.
% weisbuch1985  
This model and its variants have been studied intensely in theoretical neuroscience and statistical physics \cite{talagrand2003}, but investigations into its utility for memory and coding have mainly focused on storing collections of patterns $X$ using a ``one-shot" outer-product rule (OPR) for learning, which essentially assigns weights between neurons to be their correlation, an early idea in neuroscience \cite{hebb49}.   Independent of learning, at most $2n$ randomly-generated dense patterns can be simultaneous memories in a Hopfield network with $n$ neurons \cite{cover1965}. 

Despite this restriction, super-linear capacity in Hopfield networks is possible for special pattern classes and connectivity structures.  For instance, if patterns to memorize contain many zeroes, it is possible to store nearly a quadratic number \cite{amari1989characteristics}.
Other examples are random networks, which have $\approx 1.22^n$ memories asymptotically \cite{tanaka1980analytic}, and networks storing all permutations \cite{platt1986}. In both examples of exponential storage, however, memories have vanishingly small basins of attraction, making these networks ill-suited for noise-tolerant pattern storage.  Interestingly, the situation is even worse for networks storing permutations:  any Hopfield network storing permutations will not recover the derangements (more than a third of all permutations) from asymptotically vanishing noise (see Supplemental Material).

In this letter, we design a family of sparsely-connected $n$-node Hopfield networks with 
\begin{equation}\label{num_mem}
 \frac{2^{\sqrt{2n} + \frac{1}{4}}}{n^{1/4} \sqrt{\pi}}
\end{equation}
robust memories, asymptotically, by minimizing probability flow \cite{sohl2011}.  To our knowledge, this is the first rigorous demonstration of super-polynomial noise-tolerant storage in recurrent networks of simple linear threshold elements.  The approach also provides a normative, convex, biologically plausible learning mechanism for discovering these networks from small amounts of data and reveals new connections between binary McCulloch-Pitts neural networks, Shannon optimal error-correcting codes, and computational graph theory.

%Our networks have applications to coding theory and its relation to statistical physics \cite{vicente2002low} as they provide  a new family of nonlinear 
% $({v \choose k}, {v \choose 2}, 2k-2)$ 

\textbf{Background}.
The underlying probabilistic model of data in the Hopfield network is the non-ferromagnetic \textit{Lenz-Ising model} \cite{ising25} from statistical physics, more generally called a Markov random field in the literature, and the model distribution in a fully observable Boltzmann machine \cite{ackley1985learning} from artificial intelligence.  The states of this discrete  distribution are length $n$ binary column vectors ${\bf x} = (x_1,\ldots, x_n)$ each having probability $p_{{\bf x}}  =\frac{1}{Z} \exp \left( - E_{\mathbf{x}}   \right)$, in which $E_{\mathbf x} = -\frac{1}{2}\mathbf x^{\top} \mathbf J \mathbf x + \theta^{\top}\mathbf  x$ is the \textit{energy} of a state, $\mathbf{J} \in \mathbb R^{n \times n}$ is a symmetric matrix with zero diagonal (the
\textit{weight matrix}), the vector $\theta \in \mathbb R^n$ is a \textit{threshold} term, and $Z = \sum_{\mathbf{x}}\exp(-E_{\mathbf{x}})$ is the \textit{partition function}, the normalizing factor ensuring that $p_{\mathbf{x}}$ represents a probability distribution.  In theoretical neuroscience, rows $\mathbf{J}_e$ of the matrix $\mathbf{J}$ are interpreted as abstract ``synaptic" weights $J_{ef}$ connecting neuron $e$ to other neurons $f$.

%The Lenz-Ising model $\mathbf{p} = (p_{{\bf x}})_{\mathbf{x} \in \{0,1\}^n}$ is known to have maximum entropy over all distributions with its first- and second-order statistics \cite{CoverThomas} and often can be determined from very few of its samples \cite{geman1984stochastic, chatterjee2011random}.

%\begin{figure}[t!]
%\begin{center}
%\includegraphics[width=1.0\linewidth]{robustness_V128_S30_nT10_absp0_250_plotAllFalse.pdf}
%\end{center}
%\vspace{-.2in}
%\caption{\textbf{Exponential memories, exponential attraction basins}.  Denoising performance of Hopfield networks storing all $64$-cliques in $v=128$ vertex graphs using a dense $8128$-bit network minimizing probability flow (\ref{PFlow}) on $50000$ random cliques (light gray line), sparsely-connected $(x, 0, 1)$ network with $x$ as in (\ref{large_dev_x}) and $p=1/4$ (gray), or MPF theoretical optimum (\ref{MPF_theory_x}) (black).  $200$ cliques chosen uniformly at random were $p$-corrupted for different $p$ and then dynamics were converged initialized at noisy cliques.  The plot shows the fraction of cliques correctly recovered as a function of the pattern corruption $p$.  Dotted lines are average bits retrieved correctly.}
%\vspace{-.2in}
%\label{clique_stability_fig}
%\end{figure}

\begin{figure}[t!]
\begin{center}
\textbf{a)} \includegraphics[width=5 in]{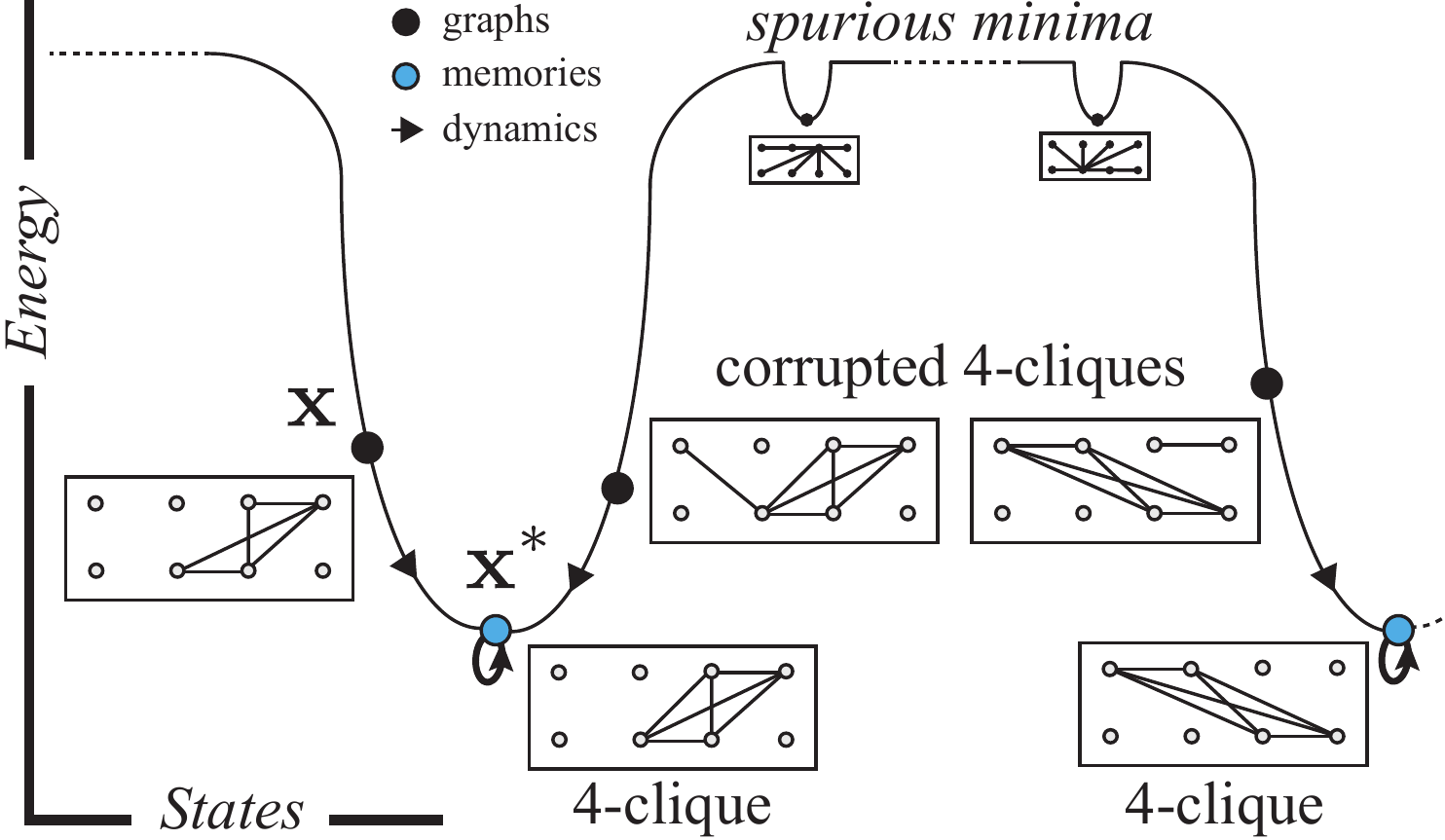} \\
\vspace{.8cm}
\textbf{b)} \includegraphics[width=5 in]{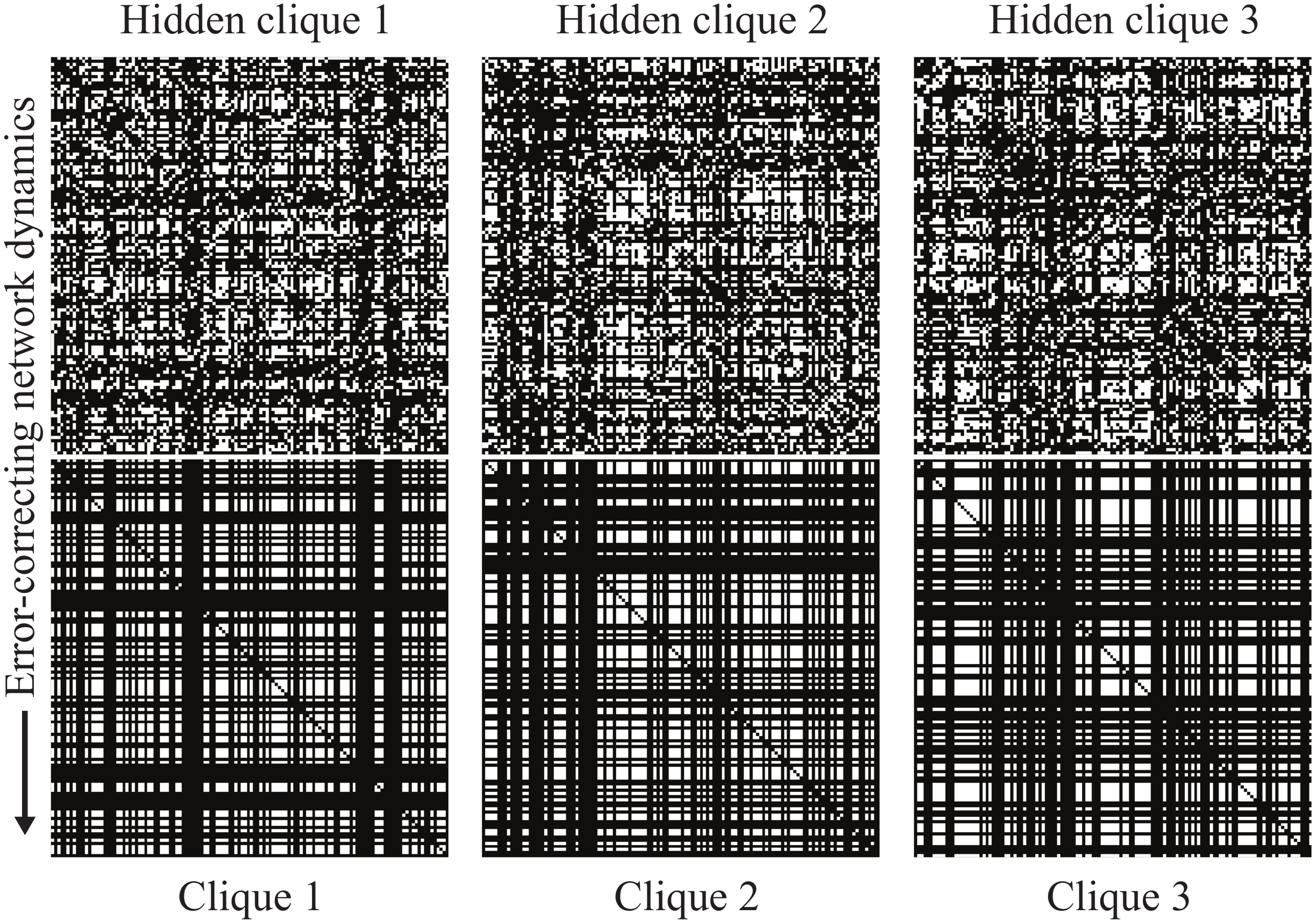} 
\caption{\textbf{a)} \textbf{Illustration of the energy landscape and discrete dynamics in a Hopfield network} having robust storage of all $4$-cliques in graphs on $8$ vertices.  The deterministic network dynamics sends three corrupted cliques to graphs with smaller energy, converging on the underlying $4$-clique attractors.
\textbf{b)} \textbf{Learning to solve $\approx 10^{37}$ ``Hidden clique" problems}. (Bottom) Adjacency matrices of three 64-cliques on $v = 128$ vertices.  (Top) Adjacency matrices of noisy versions of the cliques having, on average, 1219 bits corrupted out of $n = 8128$ from the original.  Converging dynamics of a symmetric $3$-parameter network $(x, y, z) = (.0107, 0, 1)$ with minimum probability flow initialized at these noisy cliques uncovers the originals.}
\label{hop_clique_fig_illus}
\vspace{-.8cm}
\end{center}
\end{figure}

The pair $(\mathbf{J}, \theta)$ determines an asynchronous deterministic (``zero-temperature") \textit{dynamics} on states $\mathbf x$ by replacing each $x_e$ in $\mathbf x$ with the value:
\begin{equation}\label{Hopdynamics}
x_e = \left\{\begin{array}{cc}1 & \ \ \text{if } \ \sum_{f \neq e} {J_{ef} x_f}  > \theta_e  \\ &  \\0 & \text{otherwise,} \ \ \ \ \  \end{array}\right.
\end{equation}
in a random, but fixed, order through all neurons $e = 1, \ldots, n$.  The quantity $I_e = \langle \mathbf{J}_e, \mathbf{x} \rangle$  in (\ref{Hopdynamics}) is often called the \textit{feedforward input} to neuron $e$ and may be computed by linearly combining discrete input signals from neurons with connections to $e$.  
Let $\Delta E_e$ (resp. $\Delta x_e = \pm 1, 0$) be the energy (resp. bit) change when applying  (\ref{Hopdynamics}) at neuron $e$.
The relationship $\Delta E_e = -\Delta x_e (I_e - \theta_e)$ guarantees that network dynamics does not increase energy.
Thus, each initial state $\mathbf x$ will converge in a finite number of steps to its \textit{attractor} $\mathbf x^{*}$ (also \textit{fixed-point}, \textit{memory}, or \textit{metastable state}); e.g., see Fig.~\ref{hop_clique_fig_illus}.   The biological plausibility and potential computational power of the update (\ref{Hopdynamics}) inspired early theories of neural networks \cite{mcculloch1943, rosenblatt1962principles}.

We now formalize the notion of robust memory storage for families of Hopfield networks.  The \emph{$p$-corruption} of  $\mathbf{x}$ is the random pattern $\mathbf{x}_p$ obtained by replacing each $x_e$ by $1-x_e$ with probability $p$, independently.  The $p$-corruption of a state differs from the original by $pn$ bit flips on average so that for larger $p$ it is more difficult to recover the original binary pattern; in particular, $\mathbf{x}_{\frac{1}{2}}$ is independent of $\mathbf{x}$. 
Given a Hopfield network, the memory $\mathbf{x}^\ast$ has \emph{$(1-\epsilon)$-tolerance} for a $p$-corruption if the dynamics can recover $\mathbf{x}^\ast$ from $\mathbf{x}^\ast_p$ with probability at least $1-\epsilon$. The \emph{$\alpha$-robustness} $\alpha(X, \epsilon)$ for a set of states $X$ is the most $p$-corruption every state $(1-\epsilon)$-tolerates.
%, a property both of the network and the collection of patterns $X$. 
Finally, we say that a sequence of Hopfield networks $\mathcal{H}_n$ \textit{robustly stores} states $X_n$ with robustness index $\alpha > 0$ if the following limit exists and equals $\alpha$:
\begin{equation}
\lim_{\epsilon \to 0^{+}}  \lim_{n \to \infty} \inf \left \{\alpha(X_n,\epsilon), \alpha(X_{n+1},\epsilon), \ldots \right \} = \alpha.
\end{equation}
%Formally, having a robustness index for a family of networks implies that given any desired tolerance $1-\epsilon$ around a memory $\mathbf{x}^{*}
If $\alpha$ is the robustness index of a family of networks then the chance that dynamics does not recover an $\alpha$-corrupted memory can be made as small as desired by devoting more neurons.  

%Informally, a family of networks with $n$ neurons stores its patterns $\alpha$-robustly if $\alpha n$ bit errors in a memory are nearly always recovered by converging dynamics, in the limit of large $n$.  

To determine parameters $(\mathbf{J}, \theta)$ in our Hopfield networks from a set of training patterns $X$, we minimize the following instantiation of the  \textit{probability flow} \cite{sohl2011} objective function:
\begin{equation}\label{PFlow}
%\mathcal{F}(\mathbf{J}, \theta) = 
\frac{1}{|X|} \sum_{\mb x \in X} \ \sum_{\mb x' \in \mathcal{N}(\mb x)} \exp \left(\frac{E_{\mb x}-E_{\mb x'}}{2}\right),
 \end{equation}
where $\mathcal{N}(\mb x)$ are those neighboring states $\mathbf{x}'$ differing from $\mathbf{x}$ by a single flipped bit.
It is elementary that a Hopfield network has memories $X$ if and only if the probability flow (\ref{PFlow}) can be arbitrarily close to zero, motivating the application of minimizing (\ref{PFlow}) to finding such networks  \cite{HS-DK201}.  The probability flow is convex, consists of a number of terms linear in $n$ and the size of $X$, and avoids the exponentially large partition function $Z = \sum_{\mathbf{x} \in \{0,1\}^n}\exp(-E_{\mathbf{x}})$.

\textbf{Results}.  Let $v$ be a positive integer and set $n = \frac{v(v-1)}{2}$.  A state $\mathbf{x}$ in a Hopfield network on $n$ nodes represents a simple undirected graph $G$ on $v$ vertices by interpreting a binary entry $x_{e}$ in $\mathbf{x}$ as indicating whether edge $e$ is in $G$ ($x_{e} = 1$) or not ($x_{e} = 0$).  A $k$-\textit{clique} $\mathbf{x}$ is one of the ${v \choose k} =  \frac{v \cdot (v-1)\cdots (v-k+1)}{k \cdot (k-1)\cdots 2 \cdot 1}$ graphs consisting of $k$ fully connected nodes and $v-k$ other isolated nodes.  We design Hopfield networks that have all $k$-cliques on $2k$ (or $2k-2$) vertices as robust memories.  For large $n$, the count ${2k \choose k}$ approaches (\ref{num_mem}) by Stirling's approximation.  Fig.~1\textbf{a} depicts a network with $n = 28$ neurons storing $4$-cliques in graphs on $v = 8$ vertices.

\begin{figure}
\centering
\includegraphics[width=6in]{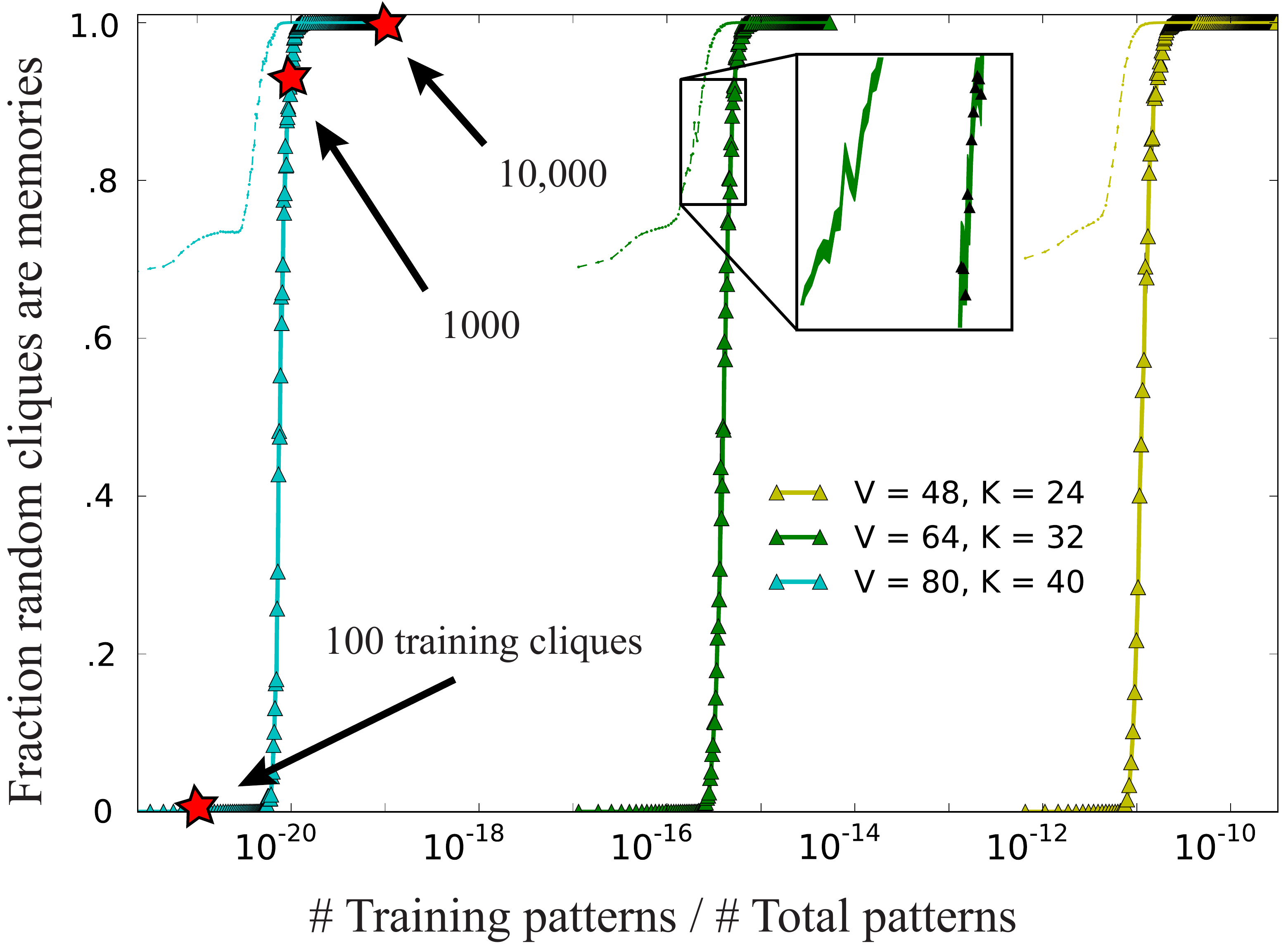}
\caption{\textbf{Learning critical networks with exponential memory by minimizing probability flow on few training patterns}.  
For numbers of vertices $v = 80, 64, 48$ ($k = 40, 32, 24$) with $50$ trials each, the average percent of 1000 randomly drawn cliques that are memories vs. the fraction of training samples to total number of $k$-cliques. Inset displays enlarged version of the region demarcated by black square; filled regions indicate standard deviation errors over these 50 trials. Dotted lines are average percentage of correct bits after converging dynamics.  
}
\vspace{-.2in}
\label{fig_small_samples}
\end{figure}

Our first result is that numerical minimization of probability flow over a vanishingly small critical number of training cliques determines linear threshold networks with exponential memory.
%In the case $z = 1$ and fixed $y < 0$, we have to optimize the $1$ variable unconstrained problem:
%\[ f(u) = \frac{au^{k-1}}{w v^{k-2}} + \frac{w}{u^{2(k-2)}v^{{k-2 \choose 2}}} + \frac{bv^{{k \choose 2}}}{w},\]
%where $u = e^x$, $v = e^y$, $w = e^z$ ($= e^1 = e$), $a = k(v-k) /  {k \choose 2}$, $b ={v-k \choose 2} / {k \choose 2}$. 
%For $v = 2k$, we should be able to explicitly solve this for $u$ which should then give us the .
We fit all-to-all connected networks on $n = 3160, 2016, 1128$ neurons ($v = 80, 64, 48$; $k=40,32, 24$) with increasing numbers of randomly generated $k$-cliques as training data by minimizing (\ref{PFlow}) using the limited-memory Broyden--Fletcher--Goldfarb--Shanno (L-BFGS) algorithm \cite{nocedal1980}.  In Fig.~\ref{fig_small_samples}, we plot the percentage of 1000 random new $k$-cliques that are memories in these networks after training as a function of the ratio of training set size to total number of $k$-cliques.  Each triangle in the figure represents the average of this fraction over 50 networks, each given the same number of randomly generated (but different) training data.
The finding is that  a critical amount of training data gives storage of all $k$-cliques.  Moreover, this count is significantly smaller than the total number of patterns to be learned.

\begin{figure}
\centering
\textbf{a)} \includegraphics[width=1.0in]{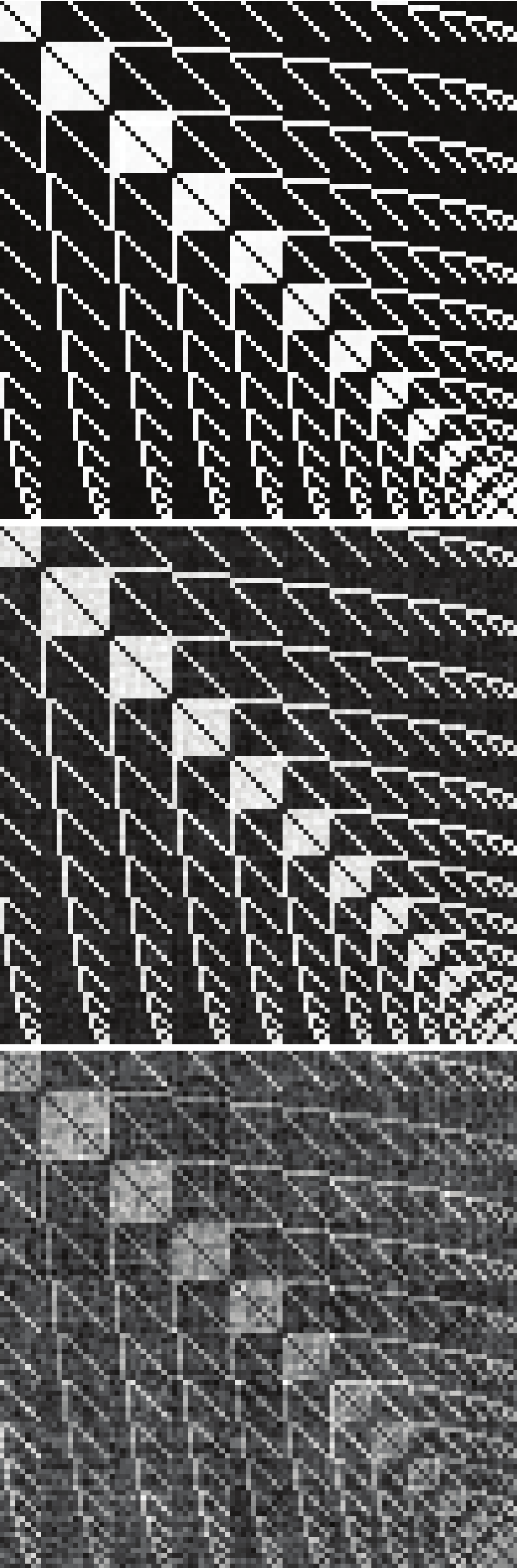}
\textbf{b)} \includegraphics[width=5.0in]{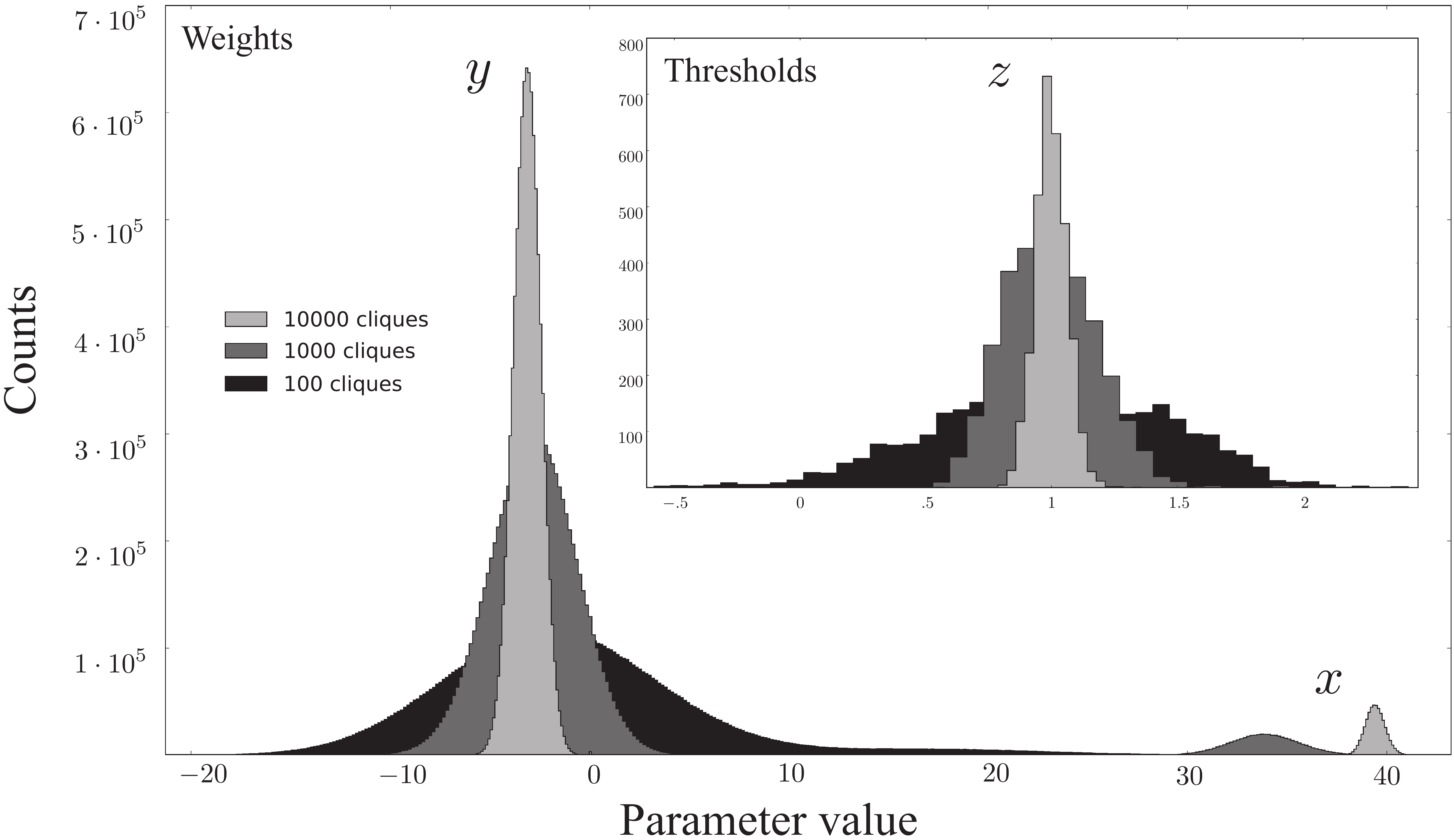}
%\textbf{c)}\includegraphics[width=4.3in]{learning_NVs_xlog3_S108_nT50_STD_ERR0_MEAN_ALPHA0.pdf}
\caption{\textbf{Distribution of network parameters learned by minimizing probability flow (MPF) sharpens around three critical values}.  
\textbf{a}) Portion of network weights $\mathbf{J}$ after minimizing (\ref{PFlow}) given 100 (bottom), 1000 (middle), or 10000 (top) random $40$-cliques $X$ (of about $10^{23}$ in total) on $v = 80$ vertices.  These networks represent the marked points in Fig.~\ref{fig_small_samples}.
\textbf{b}) Histograms of weight and threshold parameters for networks in \textbf{a} (histogram of thresholds $\theta$ in inset).  Network parameters are scaled  so that thresholds have mean 1 (this does not affect the dynamics). %Groups of similar network weights and thresholds are labelled with corresponding parameter $x$, $y$, $z$.
}
\vspace{-.2in}
\label{fig_recovery}
\end{figure}

In Fig.~\ref{fig_recovery}\textbf{a}, we display a portion of the weight matrix with minimum probability flow representing a $v = 80$ network (4,994,380 weight and threshold parameters) given $100$ ($\approx$1e-21\% of all $40$-cliques), $1000$ (1e-20\%), or $10000$ (1e-19\%) randomly generated $40$-cliques as training data; these are the three special red points marked in Fig.~\ref{fig_small_samples}.  In Fig.~\ref{fig_recovery}\textbf{b}, we plot histograms of the network parameters at these three training set sizes and find that the weights and thresholds become highly peaked and symmetric about three quantities.  Below, we shall analytically minimize the flow objective to obtain critical symmetric parameters.
 
Next, we derive a biologically plausible learning rule for adapting these network parameters.  
Given a training pattern $\mathbf{x}$, the \textit{minimum probability flow} (MPF) learning rule moves weights and thresholds in the direction of steepest descent of the  probability flow objective function (\ref{PFlow}) evaluated at $X = \{\mathbf{x}\}$. Specifically, for $e \neq f$ the rule takes the form:
\begin{equation}\label{MPF_update_rule}
\begin{split}
\Delta J_{ef} & \propto - x_f \Delta x_e \exp(-\Delta E_e/2),\\
\Delta \theta_e & \propto \Delta x_e \exp(-\Delta E_e/2).
\end{split}
\end{equation}
After learning, the weights between neurons $e$ and $f$ are symmetrized to: $\frac{1}{2}(\mathbf{J}_{ef} + \mathbf{J}_{fe})$, which preserves the energy function and guarantees that dynamics terminates in memories.  As update directions (\ref{MPF_update_rule}) descend the gradient of an infinitely differentiable convex function, learning rules based on them have good convergence rates \cite{hazan2007logarithmic}.  
Moreover, when neurons $e$ and $f$ are both active in (\ref{MPF_update_rule}),  weights increase, while when they are different they decrease, consistent with Hebb's postulate \cite{hebb49}, a basic hypothesis about neural synaptic plasticity.  In fact, approximating the exponential function with unity in (\ref{MPF_update_rule}) gives a variant of classical OPR learning.  Adaptation (\ref{MPF_update_rule}) is also \textit{local} in that updating weights between 2 neurons only requires their current state/threshold and feedforward input from nearby active neurons. 

We now analytically minimize probability flow to determine explicit networks achieving robust exponential storage.   
%We do this by deriving optimal conditions for a sparsely connected network to achieve robust exponential storage.  
To simplify matters, we first observe by a symmetrizing argument (see Supplementary Material) that there is a network storing all $k$-cliques if and only if there is one with constant threshold $\theta =  (z, \ldots, z) \in \mathbb{R}^n$ and satisfying for each pair $e \neq f$, ether $J_{ef} = x$ (whenever $e$ and $f$ share one vertex) or $J_{ef} =  y$  (when $e$ and $f$ are disjoint).  Weight matrices approximating this symmetry can be seen in Fig.~\ref{fig_recovery}\textbf{a}.  
In this case, the energy of a graph $G$ with $\#E(G)$ edges is the following linear function of $(x,y,z)$:
\begin{equation}\label{graph_energy}
E_G(x,y,z) = - x \cdot S_1(G)  -  y \cdot S_0(G)  + z \cdot \#E(G),
\end{equation}
in which $S_1(G)$ and $S_0(G)$ are the number of edge pairs in the graph $G$ with exactly one or zero shared vertices, respectively.

Consider the minimization of (\ref{PFlow}) over a training set $X$ consisting of all ${v \choose k}$ $k$-cliques on $v = 2k-2$ vertices (this simplifies the mathematics), restricting networks to our $3$-parameter family $(x,y,z)$.  When $y = 0$, these networks are sparsely-connected, having a vanishing number of connections between neurons relative to total population size.   Using single variable calculus and formula (\ref{graph_energy}), it can be shown that for any fixed positive threshold $z$, the minimum value of (\ref{PFlow}) is achieved uniquely at the parameter setting $(x,0,z)$, where
\begin{equation}\label{MPF_theory_x}
x = \frac{2z}{3k - 5}.
%x = \frac{2}{3k - 5} \left[ z  + \ln \frac{k-2}{v-k} \right].
\end{equation}
This elementary calculation gives our first main theoretical contribution.

\textbf{Theorem 1}.  \textit{McCulloch-Pitts networks minimizing probability flow can achieve robust exponential memory}.

%Rather remarkably, as we now show using a large deviation theory argument,
%\begin{equation*}
%a e^{(k-1)r - s(k-2) - t} +  e^{t - 2(k-2)r - {k -2 \choose 2}s} +  b e^{{k \choose 2}s-t},
%\end{equation*}
%where $a = k(v-k) /  {k \choose 2}$, $b ={v-k \choose 2} / {k \choose 2}$.  When $r$, $s$, and $t$ are unconstrained, then (\ref{PFlow}) has no critical points, but when $s = 0$ and $t$ is fixed, 
%

We prove Theorem 1 using the following large deviation theory argument; this approach also allows us to design networks 
achieving Shannon's bound for low-density error-correcting codes.  Fix $v = 2k$ (or $v = 2k-2$) and consider a $p$-corrupted clique.  Using Bernstein's concentration inequality for sums of bounded random variables \cite{bernstein1924}, it can be shown that an edge in the clique has $2k$ neighboring edges at least, on average, with standard deviation of order $\sqrt{k}$ (see Supplemental Material). This gives the fixed-point requirement from (\ref{Hopdynamics}):
\begin{equation*}
2kx + o(x\sqrt{k}\ln k)  > z.
\end{equation*}
On the other hand, a non-clique edge sharing a vertex with the clique has $k(1+2p)$ neighbors at most, on average, with standard deviation of  order $\sqrt{k}$.  Therefore, for a $k$-clique to be a memory, this forces again from (\ref{Hopdynamics}):
\begin{equation*}
k(1+2p)x + o(x\sqrt{k}\ln k) \leq z,
\end{equation*}
and any other edges will disappear when this holds.  (Here, we use ``little-o" notation $o(\cdot)$.) 

It follows that the optimal setting (\ref{MPF_theory_x}) for $x$ minimizing probability flow gives robust storage (with a single parallel dynamics update) of all $k$-cliques for $p < 1/4$.  This proves Theorem 1.
It is possible to do better than robustness index $\alpha = 1/4$ by setting $x = \frac{1}{2}\left[\frac{z}{2k} + \frac{z}{k(1+2p)}  \right] = \frac{z(3+2p)}{4k(1 + 2p)}$, which satisfies the above fixed-point requirements with probability approaching 1 for any fixed $p < 1/2$ and increasing $k$.
We have thus also demonstrated:

\textbf{Theorem 2}.
\textit{There is a family of Hopfield networks on $n = {2k \choose 2}$ neurons that robustly store ${2k \choose k} \approx \frac{2^{\sqrt{2n} + \frac{1}{4}}}{n^{1/4} \sqrt{\pi}}$ memories with robustness index $\alpha = 1/2$.}

\begin{figure}
\begin{center}
\includegraphics[width=6.5 in]{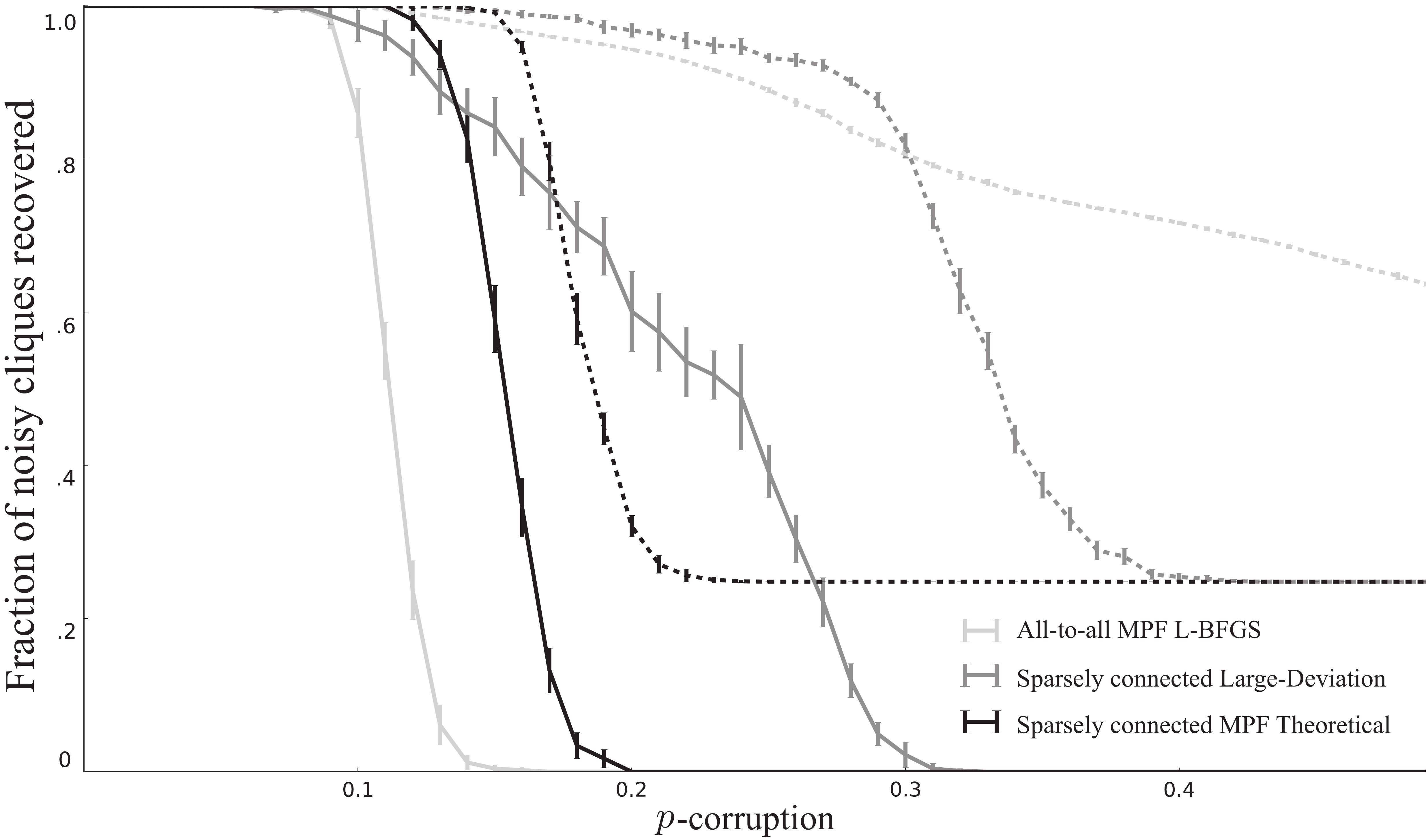}
%\textbf{b} \includegraphics[width=1.25 in]{hidden_clique.pdf} 
\caption{\textbf{Robust exponential storage in networks of McCulloch-Pitts neurons}.   Error-correction performance of Hopfield networks storing all $64$-cliques in $v=128$ vertex graphs using a fully-connected $8128$-bit network minimizing probability flow (\ref{PFlow}) on $50,000$ random $64$-cliques (light gray line), a sparsely-connected $(x, 0, 1)$ network with Large Deviation setting $x = \frac{3+2p}{4k(1 + 2p)}$ and $p=1/4$ (gray), or a sparsely-connected MPF theoretical optimum (\ref{MPF_theory_x}) (black).  Over 10 trials, one hundred $64$-cliques chosen uniformly at random were $p$-corrupted for different $p$ and then dynamics were converged initialized at noisy cliques.  The plot shows the fraction of cliques completely recovered vs. pattern corruption $p$ (standard deviation error bars).  Dotted lines are average number of bits in a pattern retrieved correctly after converging network dynamics.}
\label{hop_clique_fig}
%\vspace{-.8cm}
\end{center}
\end{figure}

In Fig.~\ref{hop_clique_fig}, we show robust storage of the ($\approx 10^{37}$) $64$-cliques in graphs on $128$ vertices using three $(x,y,z)$ parameter specializations designed here.

\textbf{Discussion}. The biologically-inspired networks introduced in this work constitute a new nonlinear error-correcting scheme that is simple to implement, parallelizable, and achieves the Shannon bound \cite{Shan48a} for low-density codes over a binary symmetric channel ($\alpha = 1/2$). 
There have been several other approaches to optimal error-correcting codes derived from a statistical physics perspective; for a comprehensive account, we refer the reader to \cite{vicente2002low}.  See also \cite{gripon2011sparse, kumar2011exponential, karbasi2013iterative, curto2013combinatorial} for related recent work on neural network architectures with large memory.  
%, notably \cite{sourlas1989}

Although we have focused on minimizing probability flow to learn parameters in our discrete neural networks, several other strategies exist.  For instance, one could maximize the (Bayesian) likelihood of cliques given network parameters, though any strategy involving a partition function over graphs will run into challenging algorithmic complexity issues \cite{jerrum1993polynomial}.  
Contrastive divergence \cite{ackley1985learning} is another popular method to estimate parameters in discrete maximum entropy models. While this approach avoids the partition function, it requires a nontrivial sampling procedure that precludes exact determination of optimal parameters.
%The flow objective is strictly convex for generic training data so that a unique minimizer

% Another observation concerning rule (\ref{MPF_update_rule}) is that the so-called Metropolis-Hastings algorithm \cite{metropolis1953equation} for sampling rejects flipping bit $x_e$ with probability $[\exp(-\Delta E_e/2)]^2 = p_{\mathbf{x}'} / p_{\mathbf{x}}$, where $\mathbf{x}'$ is $\mathbf{x}$ with $x_e$ flipped, while in (\ref{MPF_update_rule}) weights are updated in proportion to the square-root of this quantity.

In addition to classical coding theory and memory modeling in neuroscience, our networks also apply to a basic question in computational graph theory called the ``Hidden clique problem" \cite{alon1998finding}.  The essential goal of this task is to find a clique that has been hidden in a graph by adding and removing edges at random.  Phrased in this language, we have discovered discrete recurrent neural networks that learn to use their cooperative McCulloch-Pitts dynamics to solve hidden clique problems efficiently.   For example, in Fig.~1\textbf{b} we show the adjacency matrices of three corrupted $64$-cliques on $v=128$ vertices returning to their original configuration by one iteration of the network dynamics through all neurons.

%Such high coding capacities have applications in signal processing of natural stimuli; e.g., for digital image compression \cite{HMK14}.

\textbf{Acknowledgements}.
We thank Kilian Koepsell and Sarah Marzen for helpful comments that enhanced the quality of this work.
Support was provided, in part, by NSF grants IIS-0917342, IIS-1219199, IIS-1219212 (CH) and DARPA Deep Learning Program FA8650-10-C-7020 (NT). *Hillar and *Tran are listed alphabetically and contributed equally.

%\vspace{-.15in}
\bibliographystyle{apsrev4-1}
\bibliography{cliquenet}

\newpage

\begin{center}
\textbf{Supplementary Info}
\end{center}

\section{Symmetric $3$-parameter $(x,y,z)$ networks}

The first step of our construction is to exploit symmetry in the following set of linear inequalities: 
\begin{equation}\label{graph_ineq}
E_{\mathbf{c}} - E_{\mathbf{c'}} < 0,% \  \text{for each $k$-clique $\mathbf{c}$ and $\mathbf{c}' \in \mathcal{N}(\mathbf{c})$}.
\end{equation}
where $\mathbf{c}$ runs over $k$-cliques and $\mathbf{c}'$ over vectors differing from $\mathbf{c}$ by a single bit flip.
%for each $k$-clique $\mathbf{c}$ and $\mathbf{c}' \in \mathcal{N}(\mathbf{c})$.  
%Here, the \textit{neighborhood} $\mathcal{N}(\mb x)$ of $\mathbf{x}$ consists of those $n$ vectors $\mathbf{x}'$ with Hamming distance $1$ from $\mathbf{x}$.  
The space of solutions to (\ref{graph_ineq}) is the convex polyhedral cone of networks having each clique as a strict local minimum of the energy function, and thus a fixed-point of the dynamics.

The permutations $P \in P_V$ of the vertices $V$ act on a network by permuting the rows/columns of the weight matrix ($\mathbf{J} \mapsto P \mathbf{J}P^{\top}$) and  thresholds ($\theta \mapsto P \theta$), and this action on a network satisfying property (\ref{graph_ineq}) preserves that property.  Consider the average $(\mathbf{\bar{J}}, \bar{\theta})$ of a network over the group $P_V$: $\mathbf{\bar{J}} =  \frac{1}{v!}\sum_{P \in P_V}P  \mathbf{J} P^{\top}$, $\bar{\theta} = \frac{1}{v!}\sum_{P \in P_V}P  \theta$, and note that if $(\mathbf{J}, \theta)$ satisfies (\ref{graph_ineq}) then so does the highly symmetric object $(\mathbf{\bar{J}}, \bar{\theta})$.
To characterize $(\mathbf{\bar{J}}, \bar{\theta})$, observe that $P \mathbf{\bar{J}} P^{\top}= \mathbf{\bar{J}}$ and $P \bar{\theta} = \bar{\theta} \ \text{for all } P \in P_V$. 

These strong symmetries imply there are $x,y,z$ such that $\bar{\theta} =  (z, \ldots, z) \in \mathbb{R}^n$ and for each pair $e \neq f$ of all possible edges:
%\vspace{-.2in}
\begin{equation*}
\bar{J}_{ef} := \left\{ \begin{array}{ccc} x & \mbox{ if } & |e \cap f| = 1 \\ y & \mbox{ if } & |e \cap f| = 0, \end{array} \right.
%\vspace{-.15in}
\end{equation*}
where $|e \cap f|$ is the number of vertices that $e$ and $f$ share.

Our next demonstration is an exact setting for weights in these Hopfield networks.

\section{Exponential storage}

For an integer $r \geq 0$, we say that state $\mathbf{x}^\ast$ is \emph{$r$-stable} if it is an attractor for all states with Hamming distance at most $r$ from $\mathbf{x}^\ast$. Thus, if a state $\mathbf{x}^\ast$ is $r$-stably stored, the network is guaranteed to converge to $\mathbf{x}^\ast$ when exposed to any corrupted version not more than $r$ bit flips away. 

For positive integers $k$ and $r$, is there a Hopfield network on $n = \binom{2k}{2}$ nodes storing all $k$-cliques $r$-stably? We necessarily have $r \leq \lfloor k/2 \rfloor$, since $2(\lfloor k/2 \rfloor+1)$ is greater than or equal to the Hamming distance between two $k$-cliques that share a $(k-1)$-subclique. In fact, for any $k > 3$, this upper bound is achievable by a sparsely-connected three-parameter network.

\textbf{Proposition 1}.  \textit{There exists a family of three-parameter Hopfield networks with $z = 1$, $y = 0$ storing all $k$-cliques as $\lfloor k/2 \rfloor$-stable states}.

The proof relies on the following lemma, which gives the precise condition for the three-parameter Hopfield network to store $k$-cliques as $r$-stable states for fixed $r$.

\textbf{Lemma 1}.  \textit{Fix $k > 3$ and $0 \leq r < k$. The Hopfield network $(\mathbf{J}(x,y), \theta(z))$ stores all $k$-cliques as $r$-stable states if and only if the parameters $x,y,z \in \mathbb R$ satisfy
\begin{align*}
M \cdot \left[ \begin{array}{c} x \\ y \end{array} \right] < \left[ \begin{array}{c} -2 \\ -2 \\ 2 \\ 2 \end{array} \right] z,
\end{align*}
where 
$$ M = \left[ \begin{array}{ll}
4(2-k)+2r & (2-k)(k-3) \\
4(2-k) & (2-k)(k-3)-2r \\
2(k-1)+2r & (k-1)(k-2) \\
2(k-1) & (k-1)(k-2)-2r
 \end{array} \right].$$
Furthermore, a pattern within Hamming distance $r$ of a $k$-clique converges after one iteration of dynamics.}

\textit{Proof}:  For fixed $r$ and $k$-clique $\mathbf{x}$, there are $2^r$ possible patterns within Hamming distance $r$ of $\mathbf{x}$. Each of these patterns defines a pair of linear inequalities on the parameters $x,y,z$. However, only the inequalities from the following two extreme cases are active constraints. All the other inequalities are convex combinations of these.
\begin{enumerate}
	\item $r$ edges in the clique with a common node $i$ are removed.
	\item $r$ edges are added to a node $i$ not in the clique. 
\end{enumerate}
In the first case, there are two types of edges at risk of being mislabeled. The first are those of the form $ij$ for all nodes $j$ in the clique. Such an edge has $2(k-2)-r$ neighbors and ${k-2 \choose 2}$ non-neighbors. Thus, each such edge will correctly be labeled as $1$ after one network update if and only if $x$, $y$, and $z$ satisfy
\begin{equation}\label{eqn.hf:case2}
2(2k-r-4)x + (k-2)(k-3)y > 2z.
\end{equation}
The other type are those of the form $\bar{i}j$ for all nodes $\bar{i} \neq i$ in the clique, and $j$ not in the clique. Assuming $r < k-1$, such an edge has at most $k-1$ neighbors and ${k-1 \choose 2} - r$ non-neighbors. Thus, each such edge will be correctly labeled as $0$ if and only if
\begin{equation}\label{eqn.hf:case2.2}
2(k-1)x + ((k-1)(k-2)-2r)y < 2z.
\end{equation}
Rearranging equations (\ref{eqn.hf:case2}) and (\ref{eqn.hf:case2.2}) yield the first two rows of the matrix in the lemma. A similar argument applies for the second case, giving the last two inequalities.

%COMMENT: omit the derivation of the second half as it's similar. 
%keep here to ensure correctness. Can remove in future git commits
%In case $2$, consider edges $ij$ for all nodes $j$ in the clique. Each such edge has $r+k-1$ neighboring edges and ${k-1 \choose 2}$ non-neighboring edges. Thus, each such edge will be labeled as 0 after one network update if and only if $x,y$ satisfy:
%\begin{equation}\label{eqn.hf:case1}
%2(r+k-1)x + (k-1)(k-2)y < 2z.
%\end{equation}
%Now, an edge $jj'$ in the clique have $2(k-2)$ neighbors and ${k-2 \choose 2} + r$ non-neighbors. For this edge to be labeled as 1 after one network update, we need
%\begin{equation}\label{eqn.hf:case1.2}
%4(k-2)x + ((k-2)(k-3)+2r)y > 2z.
%\end{equation}

From the derivation, it follows that if a pattern is within Hamming distance $r$ of a $k$-clique, then all spurious edges are immediately deleted by case 1, all missing edges are immediately added by case 2, and thus the clique is recovered in precisely one iteration of the network dynamics.  $\blacksquare$

\textit{Proof of Proposition 1}:
The matrix inequalities in Lemma 1 define a cone in $\mathbb R^3$, and the cases $z = 1$ or $z = 0$ correspond to two separate components of this cone. For the proof of Theorem 1 in the main article, we shall use the cone with $z = 1$. We further assume $y = 0$ to achieve a sparsely-connected matrix $\mathbf{J}$. In this case, the second and fourth constraints are dominated by the first and third. Thus, we need $x$ that solves:
$$ \frac{1}{2(k-1)-r} < x < \frac{1}{k-1+r}. $$
There exists such a solution if and only if
\begin{equation}\label{eqn:feasible.x}
2(k-1)-r > k-1+r \  \Leftrightarrow \ k > 2r+1.
\end{equation}
The above equation is feasible if and only if  $r \leq \lfloor k/2 \rfloor$. 
 $\blacksquare$

\section{Proofs of Theorems 1, 2}
Fix $y = 0$ and $ z = 1$. We now tune $x$ such that asymptotically the $\alpha$-robustness of our set of Hopfield networks storing $k$-cliques tends to $1/2$ as $n \to \infty$. By symmetry, it is sufficient to prove robustness for one fixed $k$-clique $\mathbf{x}$, say, the one with vertices $\{1, 2, \ldots, k\}$. For $0 < p < 1$, let $\mathbf{x}_p$ be the $p$-corruption of $\mathbf{x}$. For each node $i \in \{1, \ldots 2k\}$, let $i_{in}, i_{out}$ denote the number of edges from $i$ to other clique and non-clique nodes, respectively. With an abuse of notation, we write $i \in \mathbf{x}$ to mean a vertex $i$ in the clique, that is, $i \in \{1, \ldots, k\}$. We need the following inequality originally due to Bernstein [24].

\textbf{Lemma 2  (Bernstein's inequality)}.
Let $S_i$ be independent Bernoulli random variables taking values $+1$ and $-1$ each with probability $1/2$.  For any $\epsilon > 0$, the following holds:
\[\mathbb P \left( \frac{1}{n} \sum_{i=1}^n S_i> \epsilon \right) \leq \exp\left( -\frac{n\epsilon^2}{2+2\epsilon/3} \right).\]

% old version
%\textbf{Lemma 2}. \textit{Let $Y$ be an $n \times n$ symmetric matrix with zero diagonal, $Y_{ij} \stackrel{i.i.d}{\sim} Bernoulli(p)$. For each $i = 1, \ldots, n$, let $Y_i = \sum_jY_{ij}$ be the $i$-th row sum. Let
%$M_n = \max_{1 \leq i \leq n}Y_i$, and $m_n = \min_{1 \leq i \leq n} Y_i$. Then for any constant $c > 0$, as $n \to \infty$, we have:
%$$\mathbb{P}(|m_n - np| > c\sqrt{n}\ln n) \to 0$$ 
%and $$\mathbb{P}(|M_n - np| > c\sqrt{n}\ln n) \to 0.$$
%In particular, $|m_n - np|, |M_n - np| = o(\sqrt{n}\ln n)$}.
%
%\textit{Proof}: Fix a constant $c > 0$. By Bernstein's inequality \cite{bernstein1924}, for each $i$ and for any $\epsilon > 0$, we have
%$$ \mathbb P(Y_i - np > n\epsilon) \leq \exp\left( -\frac{n\epsilon^2}{2+2\epsilon/3} \right). $$
%Applying a union bound with $\epsilon = \frac{c\ln n}{\sqrt{n}}$, it follows that
%\begin{align*}
%\mathbb P(\max_iY_i - np > n\epsilon) &\leq  \exp\left( -\frac{n\epsilon^2}{2+2\epsilon/3} + \ln n \right) \\
%&\leq \exp \left(-\frac{\ln ^ 2 n}{3} + \ln n\right).
%\end{align*}
%As this last bound converges to $0$ with $n \to \infty$, we have proved the claim for $M_n$. Since the distribution of $Y_i$ is symmetric about $np$, a similar inequality holds for $m_n$.  $\blacksquare$

The following fact is a fairly direct consequence of Lemma 2.

\textbf{Lemma 3}. \textit{Let $Y$ be an $n \times n$ symmetric matrix with zero diagonal, $Y_{ij} \stackrel{i.i.d}{\sim} Bernoulli(p)$. For each $i = 1, \ldots, n$, let $Y_i = \sum_jY_{ij}$ be the $i$-th row sum. Let
$M_n = \max_{1 \leq i \leq n}Y_i$, and $m_n = \min_{1 \leq i \leq n} Y_i$. Then for any constant $c > 0$, as $n \to \infty$, we have:
$$\mathbb{P}(|m_n - np| > c\sqrt{n}\ln n) \to 0$$ 
and $$\mathbb{P}(|M_n - np| > c\sqrt{n}\ln n) \to 0.$$
In particular, $|m_n - np|, |M_n - np| = o(\sqrt{n}\ln n)$}.

\textit{Proof}: Fix $c > 0$. As a direct corollary of Bernstein's inequality, for each $i$ and for any $\epsilon > 0$, we have
\[\mathbb P(Y_i - np > n\epsilon - (p + \epsilon)) \leq \exp\left( -\frac{(n-1)\epsilon^2}{2+2\epsilon/3} \right).\]
It follows that 
\[\mathbb P(Y_i - np > n\epsilon) \leq \exp\left( -\frac{n\epsilon^2}{4+4\epsilon/3} \right),\]
and thus from a union bound with $\epsilon = \frac{c\ln n}{\sqrt{n}}$, we have
\begin{align*}
\mathbb P(\max_iY_i - np > c\sqrt{n}\ln n) &\leq  \exp\left( -\frac{n\epsilon^2}{4+4\epsilon/3} + \ln n \right) \\
& \leq  \exp \left(-\frac{c^2 \ln ^ 2 n}{4 + 4c} + \ln n\right).
\end{align*}
Since this last bound converges to $0$ with $n \to \infty$, we have proved the claim for $M_n$. Since $Y_i$ is symmetric about $np$, a similar inequality holds for $m_n$.  $\blacksquare$

\textbf{Corollary 1}.
\textit{Let $M_{in} = \max_{i \in \mathbf{x}} i_{in}$, $m_{in} = \min_{i \in \mathbf{x}} i_{in}$, $M_{out} = \max_{i \not\in \mathbf{x}} i_{out}$, $m_{out} = \min_{i \not\in \mathbf{x}} i_{out}$, and $M_{between} = \max_{i \not\in \mathbf{x}} i_{in}$. Then $M_{in} - k(1-p)$, $m_{in} - k(1-p)$, $M_{out} - kp$, $m_{out} - kp$, and $M_{between} - kp$
are all of order $o(\sqrt{k}\ln k)$ as $k \to \infty$ almost surely.}

\textit{Proofs of Theorem 1, 2 (robustness)}:
Let $N(e)$ be the number of neighbors of edge $e$. For each $e$ in the clique:
$$N(e) \geq 2m_{in} + 2m_{out} \sim 2k + o(\sqrt{k}\ln k) \hspace{1em} \mbox{w.h.p}.$$
To guarantee that all edges $e$ in the clique are labeled 1 after one dynamics update, we need $x > \frac{1}{N(e)}$; that is,
\begin{equation}\label{eqn:hf.gtrsim}
x > \frac{1}{2k +  o(\sqrt{k}\ln k)}.
\end{equation}
If $f$ is an edge with exactly one clique vertex, then
\begin{align*}
N(f) & \leq M_{in} + M_{out} +2M_{between} \\
&\sim k(1+2p) + o(\sqrt{k}\ln k) \hspace{1em} \mbox{w.h.p}.
\end{align*}
To guarantee that $\mathbf{x}_f = 0$ for all such edges $f$ after one iteration of the dynamics, we need $x < \frac{1}{N(f)}$; that is,
\begin{equation}\label{eqn:hf.lesssim}
x < \frac{1}{k(1+2p) + o(\sqrt{k}\ln k)}.
\end{equation}
In particular, if $p = p(k) \sim \frac{1}{2} - k^{\delta-1/2}$ for some small $\delta \in (0, 1/2)$, then taking $x = x(k) = \frac{1}{2}(\frac{1}{k(1+2p)} + \frac{1}{2k})$ would guarantee that for large $k$, both equations (\ref{eqn:hf.gtrsim}) and (\ref{eqn:hf.lesssim}) are simultaneous satisfied. In this case, $\lim_{k\to\infty}p(k) = 1/2$, and thus the family of two-parameter Hopfield networks with $x(k)$, $y = 0$, $z = 1$ has robustness index $\alpha = 1/2$.  $\blacksquare$

\section{Clique range storage}

In this section, we give precise conditions for the existence of a Hopfield network on $\binom{v}{2}$ nodes that stores all $k$-cliques for $k$ in an interval $[m,M]$, $m \leq M \leq v$. We do not address the issue of robustness as the qualitative trade-off is clear: the more memories the network is required to store, the less robust it is. The trade-off can be analyzed by large deviation principles as in Theorem 2. 

\textbf{Theorem 3}. 
\textit{ Fix $m$ such that $3 \leq m < v$. For $M \geq m$, there exists a Hopfield network on $\binom{v}{2}$ nodes which stores all $k$-cliques in the range $[m,M]$ if and only if $M$ solves the implicit equation $x_M - x_m < 0$, where
\begin{align*}
x_m &= \frac{-(4m - \sqrt{12m^2 - 52m + 57} - 7)}{2(m^2 - m - 2)}, \\
x_M &= \frac{-(4M + \sqrt{12M^2 - 52M + 57} - 7)}{2(M^2-M-2)}.
\end{align*} }

\textit{Proof of Theorem 3}:
Fix $z = 1/2$ and $r = 0$ in Lemma 1. (We do not impose the constraint $y = 0$). Then the cone defined by the inequalities in Lemma 1 is in bijection with the polyhedron $\mathcal{I}_k \subseteq \mathbb{R}^2$ cut out by inequalities: %$R_k > 0$ and $B_k < 0$, where $R_k$ and $B_k$ are lines with equations
\begin{align*}
4(k-2)x + (k-2)(k-3)y - 1 > 0, \\
2(k-1)x + (k-1)(k-2)y - 1 < 0.
\end{align*}
Let $R_k$ be the line $4(k-2)x + (k-2)(k-3)y - 1 = 0$, and $B_k$ be the line $2(k-1)x + (k-1)(k-2)y - 1 = 0$. By symmetry, there exists a Hopfield network which stores all $k$-cliques in the range $[m,M]$ if and only if $\bigcap_{k=m}^{M}\mathcal{I}_k \neq \emptyset$. 
For a point $P \in \mathbb{R}^2$, write $x(P)$ for its $x$-coordinate. Note that for $k \geq 3$, the points $B_k \cap B_{k+1}$ lie on the following curve $Q$ implicitly parametrized by $k$: 
$$Q = \left\{\left(\frac{1}{k-1}, \frac{-1}{(k-1)(k-2)}\right): k \geq 3 \right\}.$$ 
When the polytope $\bigcap_{k=m}^{M}\mathcal{I}_k$ is nonempty, its vertices are the following points: $R_M \cap R_m, R_M \cap B_m$, $B_k \cap B_{k+1}$ for $m \leq k \leq M-1$, and the points $B_M \cap R_m$. 
This defines a nonempty convex polytope if and only if
%$$x(B_k \cap B_{k+1}) \leq x(Q \cap R_M) \leq (B_M \cap B_{M+1})$$
%for $m \leq M$, and  
%$$x(B_{m-1} \cap B_m) \leq x(Q \cap R_m) \leq x(B_m \cap B_{m+1}).$$  Therefore, $\bigcap_{k=m}^M\mathcal{I}_k \neq \emptyset$ if and only if 
$$x_M := x(Q \cap R_M) < x_m := x(Q \cap R_m).$$
Direct computation gives the formulae for $x_m, x_M$ in Theorem 3. See Fig~\ref{fig:515} for a visualization of the constraints of the feasible region.
$\blacksquare$

Fixing the number of nodes  and optimizing the range $M - m$ in Theorem 3, we obtain the following result.

% Ngoc's previous verision of Theorem 4
%\textbf{Theorem 4}.
%\textit{For large $v$, there is a Hopfield network on $n = {v \choose 2}$ nodes that stores all $\approx 2^v(1 - e^{-Cv})$ cliques of size $k$ as memories, where $k$ is in the range:
%$$ \frac{1}{D+2}v \leq k \leq \frac{3D+2}{2(D+2)}v,$$
%for constants $C \approx .002$, $D  \approx 13.93$. Moreover, this is the largest possible range of $k$ for any Hopfield network.}
%
%\textit{Proof of Theorem 4}:
%From Theorem 3, for large $m, M$ and $v$, we have the approximations $x_m \approx \frac{\sqrt{12}-4}{2m}$, $x_M \approx \frac{-\sqrt{12}-4}{2M}$. Hence $x_M - x_m < 0$ when $M \lesssim \frac{2+\sqrt{3}}{2-\sqrt{3}}m \approx 13.9282 m.$ Note that $\binom{v}{k}2^{-v}$ is the fraction of $k$-cliques in $K_v$, which is also the probability of a $Binom(v, 1/2)$ variable to equal $k$. To store the most cliques, choose $m = \left(1 + \frac{2+\sqrt{3}}{2-\sqrt{3}}\right)^{-1} \approx \frac{1}{15}v$. For large $v$, approximating the binomial distribution by a Gaussian and then using Mill's ratio to bound its tail c.d.f, we see that the proportion of cliques storable tends to  
%$$\Phi(\frac{14}{15}\sqrt{v}) - \Phi(\frac{1}{15}\sqrt{v}) = 1 - 2\Phi(\frac{1}{15}\sqrt{v}) \approx 1 - \exp(-Cv),$$
%for some constant $C \approx \frac{1}{2\cdot 15^2} \approx 0.002$. The range appearing in Theorem 4 arises from rounding fractions.
%$\blacksquare$

% The version CJH worked out 
\textbf{Theorem 4}.
\textit{For large $v$, there is a Hopfield network on $n = {v \choose 2}$ nodes that stores all $\approx 2^v(1 - e^{-Cv})$ cliques of size $k$ as memories, where $k$ is in the range:
$$ m = \frac{1}{D} v \leq k \leq v = M,$$
for constants $C \approx 0.43$, $D  \approx 13.93$. Moreover, this is the largest possible range of $k$ for any Hopfield network.}

\textit{Proof of Theorem 4}:
From Theorem 3, for large $m, M$ and $v$, we have the approximations $x_m \approx \frac{\sqrt{12}-4}{2m}$, $x_M \approx \frac{-\sqrt{12}-4}{2M}$. Hence $x_M - x_m < 0$ when $M \lesssim \frac{2+\sqrt{3}}{2-\sqrt{3}}m = Dm$. Asymptotically for large $v$, the most cliques are stored when $M = Dm$ and $[m,M]$ contains $v/2$.  Consider $m = \beta v$ so that $v \geq M = D\beta v \geq v/2$, and thus $1/D \geq \beta \geq 1/(2D)$.  Next, set $u = v/2 - m = v(1/2-\beta)$ and $w = M - v/2 = v(D\beta - 1/2)$ so that storing the most cliques becomes the problem of maximizing over admissible $\beta$ the quantity:
\[\max \{u,w\} =  \max \{v(1/2-\beta),v(D\beta-1/2)\}.\]
One can now check that $\beta = 1/D$ gives the best value, producing the range in the statement of the theorem.

Next, note that $\binom{v}{k}2^{-v}$ is the fraction of $k$-cliques in all cliques on $v$ vertices, which is also the probability of a $Binom(v, 1/2)$ variable equaling $k$. For large $v$, approximating this variable with a normal distribution and then using Mill's ratio to bound its tail c.d.f. $\Phi$, we see that the proportion of cliques storable tends to  
$$1 - \Phi \left(\frac{D-1}{D} \sqrt{v} \right) \approx 1 - \exp(-Cv),$$
for some constant $C \approx \frac{(D-1)^2}{2D^2} \approx 0.43$. 
$\blacksquare$

\section{Hopfield-Platt networks}
We prove the claim in the main text that the Hopfield-Platt network will not robustly  store \textit{derangements} (permutations without fixed-points). For large $k$, the fraction of permutations that are derangements is known to be $e^{-1} \approx 0.36$.
Fix a derangement $\sigma$ on $k$, represented as a binary vector $\mathbf{x}$ in $\{0,1\}^n$ for $n = k(k-1)$. For each ordered pair $(i,j)$, $i \neq j$, $j \neq \sigma(i)$, we construct a pattern $\mathbf{y}_{ij}$ that differs from $\mathbf{x}$ by exactly two bit flips:
\begin{enumerate}
	\item Add the edge $ij$. 
	\item Remove the edge $i\sigma(i)$. 
\end{enumerate}
There are $k(k-2)$ such pairs $(i,j)$, and thus $k(k-2)$ different patterns $\mathbf{y}_{ij}$. For each such pattern, we flip two more bits to obtain a new permutation $\mathbf{x}^{ij}$ as follows:
\begin{enumerate}
	\item Remove the edge $\sigma^{-1}(j)j$.
	\item Add the edge $\sigma^{-1}(j)\sigma(i)$.
\end{enumerate}
It is easy to see that $\mathbf{x}^{ij}$ is a permutation on $k$ letters with exactly two cycles determined by $(i,j)$. Call the set of edges modified the \emph{critical edges} of the pair $(i,j)$. Note that $\mathbf{x}^{ij}$ are all distinct and have disjoint critical edges.

Each $\mathbf{y}_{ij}$ is exactly two bit flips away from $\mathbf{x}$ and $\mathbf{x}^{ij}$, both  permutations on $k$ letters. Starting from $\mathbf{y}_{ij}$, there is no binary Hopfield network storing all permutations that always correctly recovers the original state. In other words, for a binary Hopfield network, $\mathbf{y}_{ij}$ is an \emph{indistinguishable} realization of a corrupted version of $\mathbf{x}$ and $\mathbf{x}^{ij}$.

We now prove that for each derangement $\mathbf{x}$, with probability at least $1 - (1-4p^2)^{n/2}$, its $p$-corruption $\mathbf{x}_p$ is indistinguishable from the $p$-corruption of some other permutation. This implies the statement in the main text.

For each pair $(i,j)$ as above, recall that $\mathbf{x}_p$ and $\mathbf{x}^{ij}_p$ are two random variables in $\{0,1\}^n$ obtained by flipping each edge of $\mathbf{x}$ (resp. $\mathbf{x}^{ij}$) independently with probability $p$. We construct a coupling between them as follows: define the random variable $\mathbf{x}'_p$ via
\begin{itemize}
	\item For each non-critical edge, flip this edge on $\mathbf{x}'_p$ and $\mathbf{x}^{ij}$ with the same $Bernoulli(p)$.
	\item For each critical edge, flip them on $\mathbf{x}'_p$ and $\mathbf{x}^{ij}$ with independent $Bernoulli(p)$.
\end{itemize}
Then $\mathbf{x}'_p \stackrel{d}{=} \mathbf{x}_p$ have the same distribution, and $\mathbf{x}'_p$ and $\mathbf{x}^{ij}_p$ only differ in distribution on the four critical edges. Their marginal distributions on these four edges are two discrete variables on $2^4$ states, with total variation distance $1 - 4(1-p)^2p^2$. Thus, there exists a random variable $\mathbf{x}''_p$ such that $\mathbf{x}''_p \stackrel{d}{=} \mathbf{x}'_p \stackrel{d}{=} \mathbf{x}_p$, and
$$ \mathbb{P}(\mathbf{x}''_p = \mathbf{x}^{ij}_p) = 4(1-p)^2p^2. $$

In other words, given a realization of $\mathbf{x}^{ij}_p$, with probability $4(1-p)^2p^2$, this is equal to a realization from the distribution of $\mathbf{x}_p$, and therefore no binary Hopfield network storing both $\mathbf{x}^{ij}$ and $\mathbf{x}$ can correctly recover the original state from such an input. An indistinguishable realization occurs when two of the four critical edges are flipped in a certain combination. For fixed $\mathbf{x}$, there are $k(k-2)$ such $\mathbf{x}^{ij}$ where the critical edges are disjoint. Thus, the probability of $\mathbf{x}_p$ being an indistinguishable realization from a realization of one of the $\mathbf{x}^{ij}$ is at least 
$$1 - (1 - 4(1-p)^2p^2)^{k(k-2)} > 1 - (1-4p^2)^{n/2},$$
completing the proof of the claim. $\blacksquare$

\section{Examples of clique storage robustness}

\begin{figure*}[t!]
\begin{center}
\includegraphics[width=.9 \linewidth]{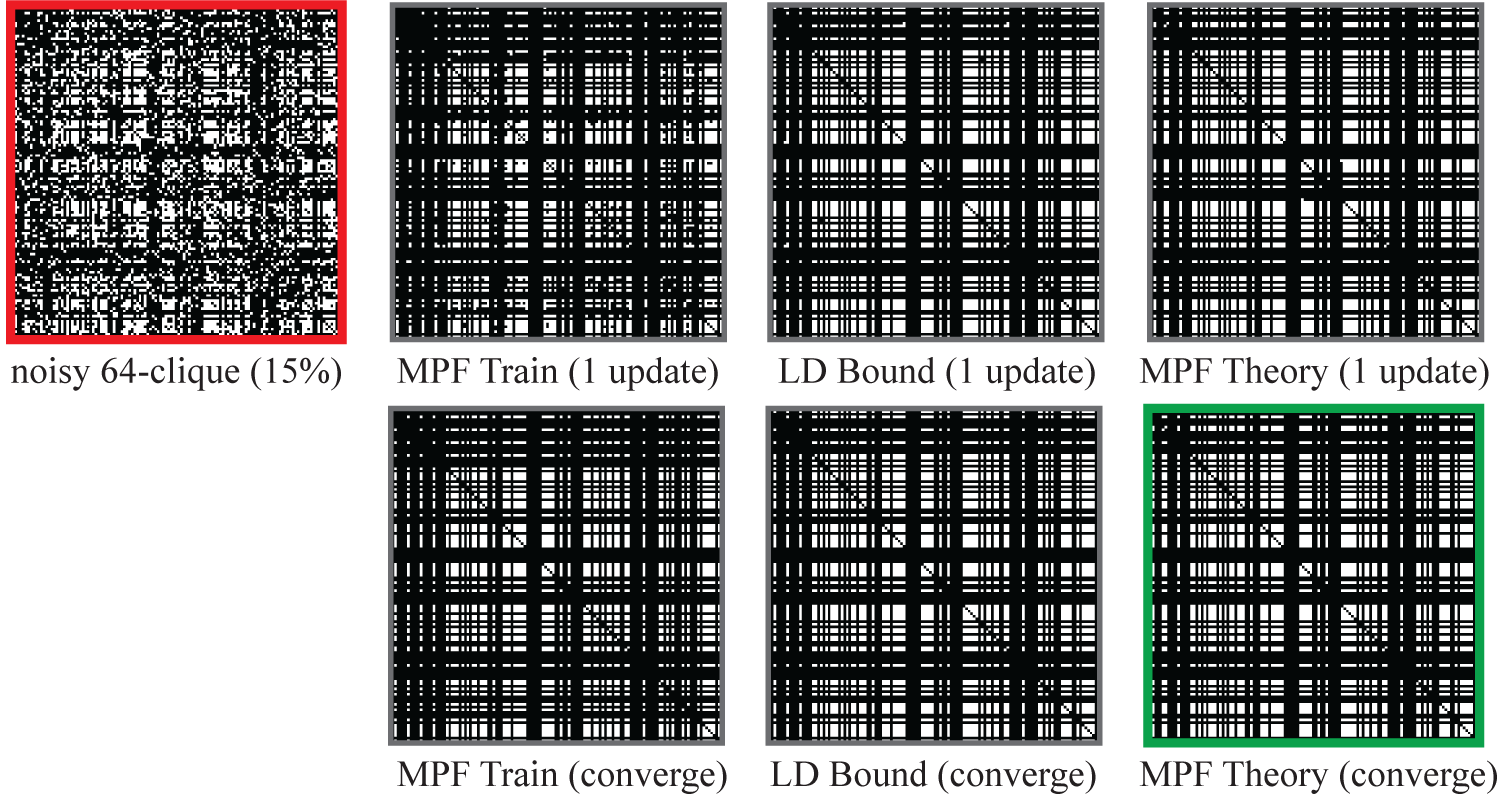} \\
\vspace{.8in}
\includegraphics[width=.9 \linewidth]{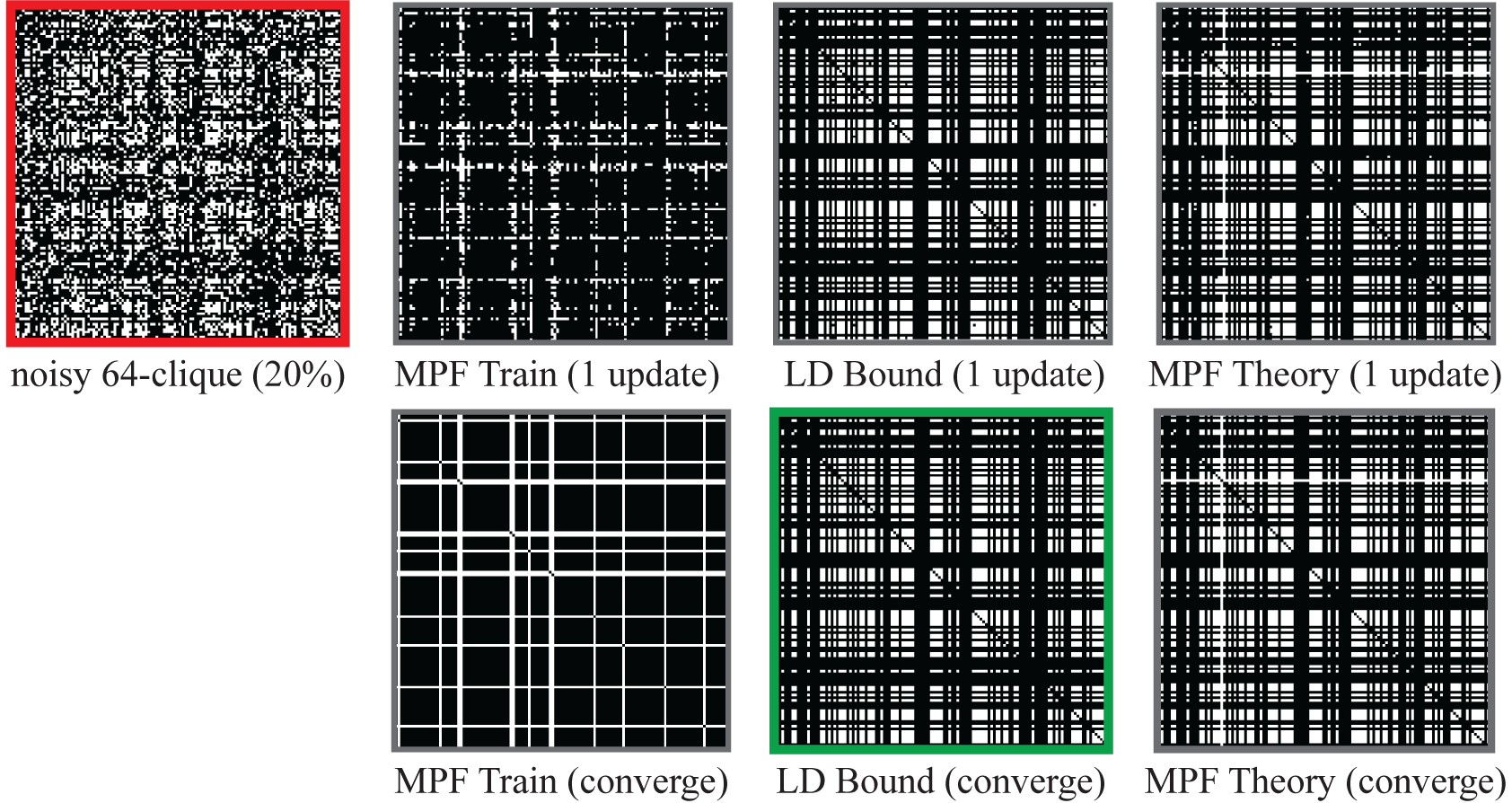}
\end{center}
\caption{\textbf{Examples of robustness for networks in Fig.~4 of main text} with $v = 128$, $k = 64$, $n = 8128$.  Adjacency matrices of noisy cliques (in red) have 1219/1625 bits corrupted out of 8128 ($p=.15 / .2$) from the original $64$-clique (in green).  Images show result of dynamics applied to these noisy patterns using networks with All-to-all MPF parameters after L-BFGS training on $50000$ $64$-cliques ($\approx$2e-31\% of all $64$-cliques), Large Deviation parameters $(x, y, z) = (.0091, 0, 1)$, or MPF Theory parameters $(x, y, z) = (.0107, 0, 1)$ from expression (7) in the main text.}
\label{clique_stability_pics}
% \vspace{-.45cm}
\end{figure*}

\begin{figure*}
	\textbf{a} \includegraphics[width=.6 \linewidth]{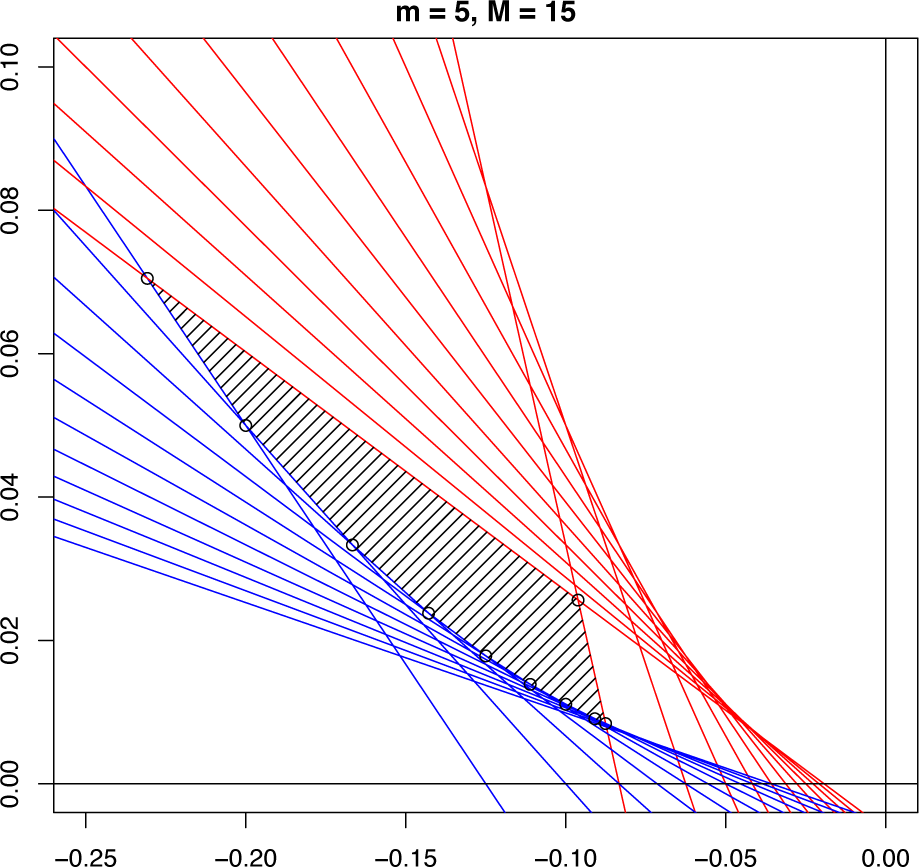} \\
\vspace{.5in}
\textbf{b} \includegraphics[width=.6 \linewidth]{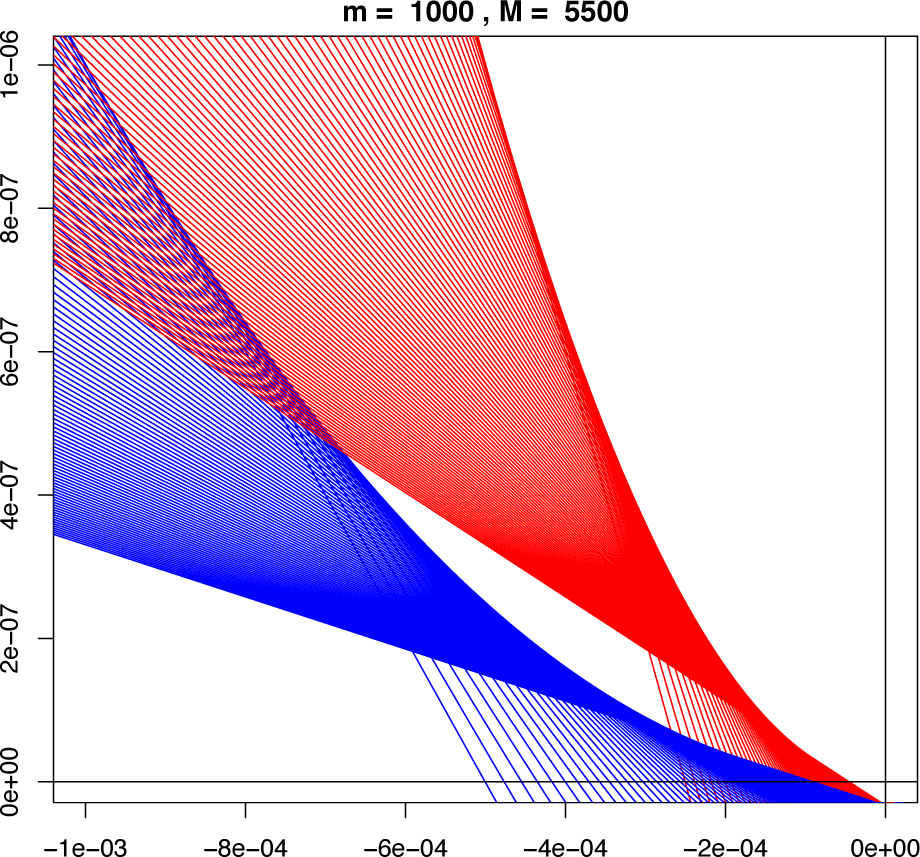}
 \caption{\textbf{Feasible exponential storage}. \textbf{a)} The shaded region is the feasible polytope   for network parameters giving clique storage for $5 \leq k \leq 15$. Black points are its vertices, the red $R_k$ and blue $B_k$ lines are linear constraints. \textbf{b)} Lines $R_k$ (red) and $B_k$ (blue) for $1000 \leq k \leq 5500$.  Note the appearance of the smooth curve $Q$ enveloping the family $B_k$ in the figure.}
\label{fig:515}
\end{figure*}

Finally, in Fig~\ref{clique_stability_pics} below we present examples of robust storage of cliques for the networks in Fig.~4 of the main text.

% \newpage

% \setcounter{figure}{4}   

% $\bigcap_{k=5}^{15}\mathcal{I}_k$

\end{document}